\documentclass[doublecol]{epl2}
\usepackage{amssymb}
\usepackage{mathrsfs}
\usepackage{graphicx}
\usepackage{float}
\usepackage{mathdots}

\title{Hidden-symmetry-protected topological phases on a one-dimensional lattice}
\author{Linhu Li\inst{1} Shu Chen\inst{1,2}}
\shortauthor{L. Li and S. Chen}

\institute{
  \inst{1} Beijing National Laboratory for Condensed Matter
Physics, Institute of Physics, Chinese Academy of Sciences, Beijing
100190, China\\
  \inst{2} Collaborative Innovation Center of Quantum Matter,
Beijing, China
}

\abstract{We demonstrate the existence of topologically nontrivial phase in a one-dimensional fermionic lattice system
subjected to synthetic gauge fields, which is beyond the standard Altland-Zirnbauer
classification of topological insulators. The
topological phase can be characterized by the presence of
degenerate zero-mode edge states or a quantized Berry phase of the
occupied Bloch band. By analyzing symmetries of the system, we
identify that the topological phase and zero-mode edge states are protected by two hidden symmetries. An
extended model with hidden symmetry breaking is also studied in
order to reveal the effect of hidden symmetries on the symmetry
protected topological phase.}

\begin{document}

\maketitle

\section{Introduction}
Exploring nontrivial topological states has attracted wide interest
in various fields of physics as stimulated by the rapid progress in
the study of topological insulators
\cite{review}. Due to their relatively simple
geometrical structures and good tunability, one-dimensional (1D)
systems with topologically nontrivial phases have attracted intense
recent studies \cite{SSH,Kitaev,Guo,Vicent,XJ_Liu,Shaolong,Ojanen,Ganeshan,Longhi,Shu,Zhu,MF1,MF2,op1,cold1,Poli,DIII,DIII2,WangZD}
 with a series of experimental progress having being
made in hybrid superconductor-semiconductor wires \cite{MF1,MF2},
photonic crystals \cite{op1} , cold atomic gases \cite{cold1}
and microwave settings \cite{Poli}. Depending on its global discrete
symmetries, such as time-reversal symmetry, particle-hole symmetry
and chiral symmetry, a 1D fermion system can be classified into ten
different symmetry classes \cite{AZ}, and five of them support
topological states \cite{symmetry_class1,tenfold,symmetry_class2}. 
%A celebrated example is the model originally proposed by Su, Schrieffer, and Heeger (SSH) to describe 1D polyacetylene \cite{SSH}, which belongs to the BDI symmetry class. With the help of state-of-art technologies of manipulating cold atoms, the SSH model has recently been experimentally realized in a 1D dimerized optical lattice and the existence of topological phase is verified by directly measuring the geometric Zak phase of the Bloch band \cite{cold1}.  
Typical 1D topological models, such as the SSH model (class BDI)
\cite{SSH} and Kitaev's Majorana chain model (class D) \cite{Kitaev}, belong to the standard ten-fold symmetry class \cite{symmetry_class1,tenfold}. A model of symmetry class AIII has been proposed for cold fermions subjected to artificial gauge fields on a 1D optical lattice \cite{XJ_Liu}. 1D topological models in class DIII \cite{DIII,DIII2,WangZD} and class CII \cite{WangZD} have also been studied very recently.

While there exist only five topologically nontrivial classes in the
standard Altland-Zirnbauer classification
\cite{AZ,symmetry_class1,tenfold,symmetry_class2}, topological
classification has been enriched when some other symmetries, e.g.,
the reflection symmetry \cite{class_reflection} and the inversion
symmetry \cite{inversion1,inversion2}, are considered in
addition of the time-reversal, particle-hole and chiral symmetries.
In principle, for every discrete symmetry, there exist corresponding
topological insulating phases with distinct physical properties
related to the specific symmetry. Usually, these topologically
nontrivial states protected by specific symmetries belong to the
category of the symmetry protected topological (SPT) state, which
can be defined as a state with trivial bulk spectrum, but nontrivial
boundary spectrum when and only when the system including the
boundary preserves the same specific symmetries
\cite{spt1,spt2,spt3}.

As most of the studied models
\cite{SSH,Kitaev,Guo,Vicent,XJ_Liu,Shaolong,Ojanen,Ganeshan,Longhi,DIII,DIII2,WangZD} can be
classified into tenfold Altland-Zirnbauer classes, it is interesting
to search for topologically nontrivial models beyond the standard
classification. In this work, we present a 1D topological model, 
which can not be classified into the standard ten-fold classes but is protected by some unusual hidden symmetries. By tuning parameters
of the model, the system can be a conductor, a trivial insulator or
a topological insulator characterized by the quantized Berry phase
\cite{Berry1,Berry2,zak}. The topologically non-trivial phase
supports doubly degenerate zero-mode edge states, which are
protected by both a hidden chiral symmetry and a hidden combination
symmetry of the
inversion and complex conjugation. By adding an additional spin-flip
term, which breaks the hidden chiral symmetry, the zero-mode edge
states no longer exist, but the model can still support
topologically nontrivial phase with degenerate edge states protected
by the hidden combination symmetry. In this case, the model can be
viewed as an extended version of the Creutz ladder model, as it can
be mapped to the Creutz ladder model \cite{Creutz} at
some special parameter regimes.

\section{Model}
We consider a tight-binding model of two-component fermionic atoms
loaded in a 1D lattice. The atoms are subjected to a synthetic gauge
potential and a Zeeman field $M_z$, which can be described by the
tight-binding Hamiltonian:
\begin{eqnarray}
H_1=&\sum_{n,ss'}\hat{c}_{n,s}^{\dagger}U_{ss'}\hat{c}_{n+1,s'}+h.c.\nonumber\\
&+M_z(\hat{c}_{n,\uparrow}^{\dagger}\hat{c}_{n,\uparrow}-\hat{c}_{n,\downarrow}^{\dagger}\hat{c}_{n,\downarrow}),\label{H1}
\end{eqnarray}
where $s$ labels the two components (up and down arrows) of the
fermions and the tunneling matrix can be written as:
\begin{eqnarray}
U=\left(
\begin{array}{cc}
-t e^{i\theta} & t_s \\
t_s & - te^{-i\theta}
\end{array}%
\right).
\end{eqnarray}
%Explicitly, the Hamiltonian reads:
%\begin{eqnarray}
%H&=& -t \sum_{n=1}^L(e^{i\theta}\hat{c}_{n,\uparrow}^{\dagger}\hat{c}_{n+1,\uparrow}+
%e^{-i\theta}\hat{c}_{n,\downarrow}^{\dagger}\hat{c}_{n+1,\downarrow})+h.c.\nonumber\\
%&&+t_s \sum_{n=1}^L
%(\hat{c}_{n,\uparrow}^{\dagger}\hat{c}_{n+1,\downarrow}+\hat{c}_{n,\downarrow}^{\dagger} \hat{c}_{n+1,\uparrow})+h.c.\nonumber\\
%&&+M_z\sum_{n=1}^L(\hat{c}_{n,\uparrow}^{\dagger}\hat{c}_{n,\uparrow}-\hat{c}_{n,\downarrow}^{\dagger}\hat{c}_{n,\downarrow}) \label{H1}
%\end{eqnarray}
%with $L$ the number of lattice sites.
For convenience, we set $t=1$ as the energy unit in the following
context. This model may be realized in two-component fermion systems with
artificial gauge potentials.

For the system under the period boundary condition (PBC), its energy
spectrum is easily calculated through a Fourier transformation
$\hat{c}_{n,s}=1/\sqrt{L}\sum_k e^{ikn}\hat{c}_{k,s}$ with $L$ the
number of lattice sites. In the momentum space, the Hamiltonian can
be written as
\begin{eqnarray}
H_1=\sum_{k} \hat{\psi}_k^{\dagger}h(k)\hat{\psi}_k,
\end{eqnarray}
where
$\hat{\psi}_k^{\dagger}=(\hat{c}_{k,\uparrow}^{\dagger},\hat{c}_{k,\downarrow}^{\dagger})$
%and
%\begin{eqnarray}
%h(k)= \left(
%\begin{array}{cc}
%M_z-2\cos{(\theta+k)} & 2t_s\cos{k} \\
%2t_s\cos{k} &
%-M_z-2\cos{(\theta-k)}%
%\end{array}%
%\right),
%\end{eqnarray}
and $ h(k)= d_0(k)I+d(k)\cdot\sigma $ with $I$ the identity matrix,
and $\sigma=(\sigma_x,\sigma_y,\sigma_z)$ the Pauli matrices acting
on the spin vector $\hat{\psi}_k$, $d_0(k)=-2\cos{\theta}\cos{k}$,
$d_x(k)=2t_s\cos{k}$, $d_y(k)=0$, $d_z(k)=2\sin{\theta}\sin{k}+M_z$.
The eigen-energies are
\begin{eqnarray}
E_{\pm}(k)=d_0(k)\pm\sqrt{d_x^2(k)+d_z^2(k)}.\label{eigenenergy}
\end{eqnarray}

\begin{figure}
\includegraphics[width=1\linewidth]{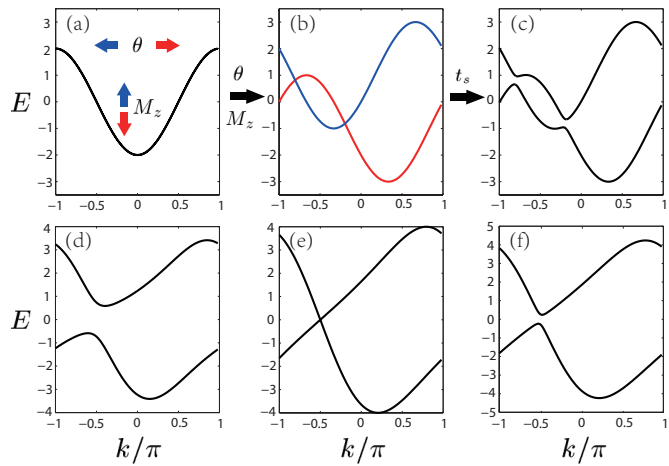}
\caption{(Color online) The spectrum in the momentum space. (a)
$M_z=t_s=\theta=0$; (b) $M_z=1,~\theta=\pi/3,~t_s=0$; (c)
$M_z=1,~\theta=\pi/3,~t_s=0.1$; (d)-(f) $\theta=\pi/3,~t_s=1$,
$M_z=1$, $\sqrt{3}$, and $2$, respectively. (a)-(c) show the
different effect of each parameter, and (d)-(f) show the variance of
spectrum versus $M_z$, which indicates the existence of a phase
transition induced by varying $M_z$.} \label{fig1}
\end{figure}

In the absence of $t_s$, the two components of fermions are
independent from each other. A nonzero $\theta$ shifts the two bands
to the left and the right respectively in the Brillouin zone, and
$M_z$ moves the two bands up and down, as showed in
Fig.\ref{fig1}(a) and (b). Further increasing the value of $M_z$
will separate the two bands completely by opening a gap between
them, and the system becomes a trivial insulator. Adding a spin-flip
term of $t_s$ will mix bands of different components of fermions. As
shown in Fig.\ref{fig1}(c), a term of $t_s=0.1$ lifts the degeneracy
of crossing points of bands in Fig.\ref{fig1}(b). When $t_s=1$, the
up and down bands are completely separated by a gap at $k=-\pi/2$
(Fig.\ref{fig1}(d)). In Fig.\ref{fig1}(d)-(f), we display the
spectrum with $\theta=\pi/3$, $t_s=1$ and different $M_z$. It is
shown that the gap is closed at $M_z=\sqrt{3}$ and reopened with
further increasing $M_z$, which indicates that a quantum phase
transition can be induced by varying $M_z$ with the transition point
determined by the gap closing point.

\begin{figure}
\includegraphics[width=0.9\linewidth]{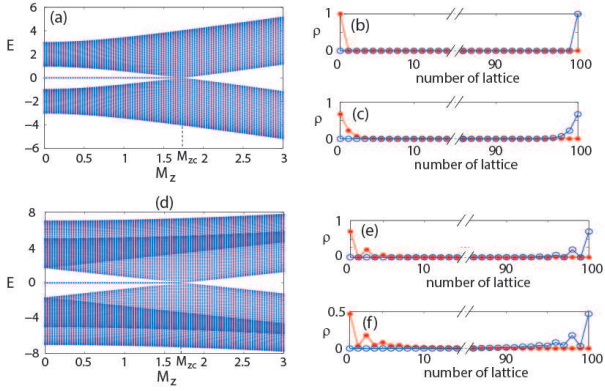}
\caption{(Color online) (a) and (d) show spectra of two different
systems under the OBC versus $M_z$, and figures on the right show
density distributions $\rho$ of zero modes for corresponding systems
on the left of figures. The parameters are: (a)-(c), $t_s=1$ and
$\theta=\pi/3$. Particularly, we have $M_z=0$ for (b) and $M_z=1$
for (c); (d)-(f) $t_s=3$ and $\theta=\pi/3$. For (e) and (f), we
have $M_z=0$ and $M_z=1$, respectively. The phase transition points
in (a) and (d) are both $M_{zc}\approx1.73$.} \label{fig2}
\end{figure}

\section{Edge states}
Next we unveil that the phase transition induced by $M_z$ is a
topological phase transition. For 1D systems, a hallmark of the
topological phase is the emergence of degenerate edge states under
the open boundary condition (OBC). To determine whether the phase
transition is topological or not, we diagonalize the Hamiltonian
under the OBC, and illustrate its energy spectrum. Defining the
single particle state as $\Psi=\sum_n^L
(\phi_{n,\uparrow}\hat{c}^{\dagger}_{n,\uparrow}+\phi_{n,\downarrow}\hat{c}^{\dagger}_{n,\downarrow})|0\rangle$,
from $H\Psi=E\Psi$, we have the eigenequations:
\begin{eqnarray}
E\phi_{n,\uparrow}=&-&[e^{i\theta}\phi_{n+1,\uparrow}+e^{-i\theta}\phi_{n-1,\uparrow}]\nonumber\\
&+&t_s[\phi_{n+1,\downarrow}+\phi_{n-1,\downarrow}]+M_z\phi_{n,\uparrow}, \nonumber\\
E\phi_{n,\downarrow}=&-&[e^{-i\theta}\phi_{n+1,\downarrow}+e^{i\theta}\phi_{n-1,\downarrow}]\nonumber\\
&+&t_s[\phi_{n+1,\uparrow}+\phi_{n-1,\uparrow}]-M_z\phi_{n,\downarrow}.
\label{eigen}
\end{eqnarray}
Numerically solving Eqs.(\ref{eigen}) under the OBC of
$\phi_{0,s}=\phi_{L+1,s}=0$, we can obtain the energy spectrum of
the system. In Fig.\ref{fig2}(a), we display the energy spectra for
the open chain with $\theta=\pi/3$ , $t_s=1$ and $L=100$ as a
function of $M_z$. It is clear that the gap closed at
$M_{zc}\approx1.73$ and reopened when $M_z>M_{zc}$, which is
consistent with the spectra under PBC as shown in
Fig.\ref{fig1}(d)-(f). The novel phenomena here is the emergence of
zero mode states for the system under OBC in the regime of $M_z <
M_{zc}$. As shown in Fig.\ref{fig2}(b)-(c), the particle density
distributions $\rho_n=|\phi_{n,\uparrow}|^2+|\phi_{n,\downarrow}|^2$
of the doubly degenerate zero mode states indicate that they are
edge states localized at the different end of the chain. The
existence or absence of zero modes in the regime of $M_z<M_{zc}$ or
$M_z>M_{zc}$ suggests that the phase transition occurring at
$M_{zc}$ is topologically nontrivial.

Next we show that these edge states can be achieved analytically in
some special parameter regions. To see it clearly, first we
reorganize the wave amplitudes $\phi_{n,s}$ as
\begin{eqnarray}
f_n=\phi_{n,\downarrow}-e^{i\beta}\phi_{n,\uparrow},~g_n=\phi_{n,\uparrow}-e^{i\beta}\phi_{n,\downarrow}
,
\end{eqnarray}
with
$e^{i\beta}=\frac{\cos{\theta}}{t_s}+i\sqrt{1-\frac{\cos^2{\theta}}{t_s^2}}$.
With these new defined functions, requiring the eigen-energy $E=0$,
Eqs. (\ref{eigen}) can be decoupled as
\begin{eqnarray}
-M_zf_n+i\mu_{+} f_{n+1}+i\mu_{-} f_{n-1}=0 , \nonumber\\
M_zg_n+i\mu_{-} g_{n+1}+i\mu_{+} g_{n-1}=0\label{decouple}
\end{eqnarray}
with $\mu_+=\sqrt{t_s^2-\cos^2{\theta}}+\sin{\theta}$,
$\mu_-=\sqrt{t_s^2-\cos^2{\theta}}-\sin{\theta}$, and $n=1,2,...,L$.
In terms of the transfer matrix form, $f_n$ and $g_n$ can be written
as
\begin{eqnarray}
\left(
\begin{array}{c}
f_{n+1}\\
f_{n}
\end{array}\right)=F\left(
\begin{array}{c}
f_{n}\\
f_{n-1}
\end{array}\right),~
\left(
\begin{array}{c}
g_{n-1}\\
g_{n}
\end{array}\right)=G\left(
\begin{array}{c}
g_{n}\\
g_{n+1}
\end{array}\right), \label{TME}
\end{eqnarray}
where
\begin{eqnarray}
F=\left(
\begin{array}{cc}
\frac{M_z}{i\mu_+} & -\frac{\mu_-}{\mu_+} \\
1 & 0
\end{array}\right),~
G=\left(
\begin{array}{cc}
\frac{-M_z}{i\mu_+} & -\frac{\mu_-}{\mu_+} \\
1 & 0
\end{array}\right)\label{transfer}.
\end{eqnarray}
%For the sake of simplicity of calculation, our definition of $G$
%transfers from $n=L$ to $n=1$, which is different from the usual
%definition of the transfer matrix $F$.
Without loss of generality,
we shall focus on the regime of $\theta \in (0,\pi)$ with
$\sin{\theta}>0$ in the following calculation.

%In the simplest case with $M_z=0$ and $t_s=\pm1$, the only non-zero
%elements of the transfer matrices are $F_{21}=G_{21}=1$, and the
%equations yield $f_1=g_L=1$, with all the other amplitudes
%$f_n=g_n=0$, as shown in Fig.\ref{fig2}(b).

For $t_s=\pm1$ and general $M_z$, the corresponding equations give solutions of
$f_n=(\frac{M_z}{2i\sin{\theta}})^{n-1}$ and $g_n=(\frac{M_z}{2i\sin{\theta}})^{L-n}$ When $|M_z|<2\sin{\theta}$
and $L\gg1$, which decrease exponentially from one end to another.
%The density distributions of zero-mode states for the example system
%with $t_s=1$, $M_z=1$ and $\theta=\pi/3$ are given in
%Fig.\ref{fig2}(c).
For the case with $M_z=0$ and general $t_s$, the solutions have the
forms of $f_{2m-1}=(-\frac{\mu_-}{\mu_+})^{m-1}$ and
$g_{2m}=(-\frac{\mu_-}{\mu_+})^{\frac{L}{2}-m}$, with
$f_{2m}=g_{2m-1}=0$ and $m=1,2,...,{L}/{2}$ for even number of
lattice sites $L$.
%as shown in Fig.\ref{fig2}(e) for the example system with $t_s=3$, $M_z=0$ and $\theta=\pi/3$.
For odd $L$, these two zero-mode solutions have
expressions similar to the ones of even case, but both $f_n$ and
$g_n$ have non-zero values only when $n=2m-1$, with
$m=1,2,...,(L+1)/2$. These edge modes also decay exponentially, and
exist only when $L\gg1$ and $|t_s|>|\cos{\theta}|$. The density distributions of zero-mode states of these two cases are given in Fig.\ref{fig2}(c) and (e) respectively.

For a general case with $M_z\neq0$ and $t_s\neq\pm1$, the edge modes
can be obtained by numerically solving Eqs.(\ref{decouple}) as shown
in Fig.\ref{fig2}(f) for the example system with $t_s=3$, $M_z=1$
and $\theta=\pi/3$. Although the analytical solution for a general
case is not accessible, we can determine the condition for the
appearance of edge states by analyzing eigenvalues of the transfer
matrices in the scheme of the transfer matrix method, which has been
widely used in 1D systems \cite{TM}. Assuming
$F(a_{f},b_{f})^T = \epsilon_f (a_{f},b_{f})^T $, the vector
$(f_{1},f_{0})^T=(f_{1},0)^T$ can be always written as a linear
superposition of the two eigenvectors of $F$, i.e.,
$(f_{1},0)^T=s_1(a_{f,1},b_{f,1})^T+s_2(a_{f,2},b_{f,2})^T$, with
$s_1$ and $s_2$ the superposition coefficients. Thus from Eq.
(\ref{TME}) we have $(f_{n+1},f_{n})^T = F^n (f_{1},f_{0})^T$, i.e.,
\begin{eqnarray}
\left(
\begin{array}{c}
f_{n+1} \\
f_{n}
\end{array}\right) =  s_1\epsilon_{f,1}^n\left(
\begin{array}{c}
a_{f,1} \\
b_{f,1}
\end{array}\right)+s_2\epsilon_{f,2}^n\left(
\begin{array}{c}
a_{f,2} \\
b_{f,2}
\end{array}\right) ,
\end{eqnarray}
which suggests that $f_n=0$ as $n\rightarrow\infty$,  i.e., the
eigenstate localized at the edge of $f_1$, if the modulus of each
eigenvalue of $F$ are smaller than unity. By solving the
characteristic equation of $F$, we can achieve $\epsilon_{f,1}=\pm
i\mu_-/\mu_+$ and $\epsilon_{f,2}=\mp i$ while
$M_z=\pm2\sin{\theta}$. Noticing $|\epsilon_{f,1}|<1$ when
$|t_s|>|\cos{\theta}|$ and $|\epsilon_{f,2}|=1$ with this specific
$M_z$, we can infer that $M_z=\pm2\sin{\theta}$ are transition
points, which separate regimes with and without edge states. With
further analyzing of the characteristic equation, we find that
$|\epsilon_{f,1}|$ decreases with the increase of $|M_z|$, whereas
$|\epsilon_{f,2}|$ increases with the increase of $|M_z|$. Hence
$|\epsilon_{f,2}|$ will exceed unity when $|M_z|>2\sin{\theta}$,
while the maximum of $|\epsilon_{f,1}|$ is
$|\epsilon_{f,1}(M_z=0)|=|\sqrt{\mu_-/\mu_+}|<1$. Thus we have the
conclusion that the zero-mode edge states $f$ only exists when
$|M_z|<2\sin{\theta}$. Similar discussion can be applied to $g_n$
and $G$ as well, while it is localized at the different end from
$f_n$. The above discussion shows that as long as
$|t_s|>|\cos{\theta}|$, topological phase transition points
$M_{zc}=\pm2\sin{\theta}$ are irrelevant to $t_s$. This conclusion
is also supported by the numerical results in Fig.\ref{fig2}(a) and
(d).
\begin{figure}
\includegraphics[width=0.9\linewidth]{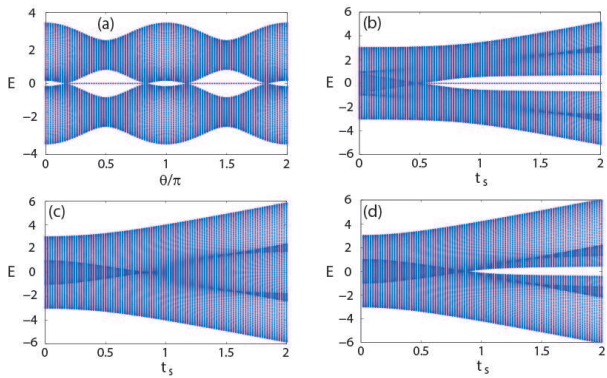}
\caption{(Color online) The OBC spectrum versus different
parameters. (a) $M_z=1$, $t_s=1$; (b) $M_z=1$, $\theta=\pi/3$; (c)
$M_z=1$, $\theta=\pi/6$; (d) $M_z=1$, $\theta=\pi/12$.} \label{fig3}
\end{figure}

The condition $|M_z|<2\sin{\theta}$ also indicates that no zero-mode
edge states could exist if $\theta=0$ and $\pi$. Although the above
discussion is limited in the regime of $\theta \in (0,\pi)$, for the
case in the regime of  $\theta \in (\pi,2\pi)$ with
$\sin{\theta}<0$, following similar procedures we can achieve the
condition for the existence of zero-mode edge modes given by
$|M_z|<-2\sin{\theta}$, and the two edge modes $f$ and $g$ localized
at the other ends opposite to the ones above. These conditions
suggest that the topological properties of this model is similar for
$\theta$ and $-\theta$. As illustrated in Fig.\ref{fig3}(a), the
spectrum under OBC versus $\theta$ is symmetric about $\theta=0$,
and varying $\theta$ while fixing other parameters will also induce
a topological phase transition with the topologically nontrivial
(trivial) phase characterized by the presence (absence) of
degenerate zero modes. In Fig.\ref{fig3}(b)-(d), we also display
spectra of systems under the OBC versus $t_s$ by fixing $M_z=1$ and
$\theta$, with $\theta=\pi/3$, $\pi/6$ and $\pi/12$, respectively.
While there always exists a conductor phase for different $\theta$
in the regime of $|t_s|<|\cos{\theta}|$,  there exist different
phases in the regime of $|t_s|>|\cos{\theta}|$ depending on whether
$\sin{\theta}$ is larger or smaller than $1/2$. We will further
discuss the phase boundary condition in the next section.

\section{Topological invariant and phase diagram}
The existence of zero-mode edge states is attributed to the
nontrivial topology of the corresponding bulk system, which can be
characterized by the Berry phase of the occupied Bloch band. In the
momentum space, the Berry phase is defined as
\begin{equation}
\gamma=\int_{-\pi}^{\pi}dk\langle u_k|i\partial_k|u_k\rangle,
\label{zak}
\end{equation}
where $u_k$ denote the occupied Bloch states which are eigenstates
of the Hamiltonian $h(k)$. In general, the Berry phase $\gamma$
across the Brillouin zone is also referred as Zak phase \cite{zak}.
For the half-filled system, the lower Bloch state is fully filled
when the two bands of $h(k)$ are completely separated. The lower
eigenstate of $h(k)$ is
\begin{eqnarray}
|u_k\rangle = \frac{1}{\sqrt{2}}\left(
\begin{array}{c}
sgn(n_x)\sqrt{1-n_z}\\
-\sqrt{1+n_z}
\end{array}%
\right), \label{uk}
\end{eqnarray}
where $n_x=d_x(k)/|d(k)|$, $n_z=d_z(k)/|d(k)|$ and
$|d(k)|=\sqrt{d_x^2+d_z^2}$. By substituting (\ref{uk}) into
(\ref{zak}), after some algebras one can obtain $ \gamma
=\frac{\pi }{2}[sgn(M_z-2\sin{\theta})-sgn(M_z+2\sin{\theta})] $
with a modulus $2\pi$, which indicates $\gamma=\pi$ when
$|M_z|<|2\sin{\theta}|$ and $\gamma=0$ when $|M_z|>|2\sin{\theta}|$.

\begin{figure}
\includegraphics[width=0.9\linewidth]{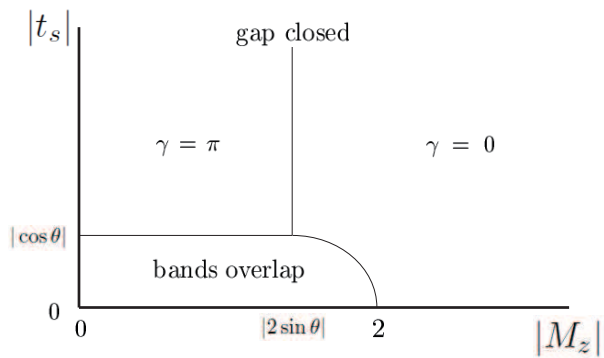}
\caption{The phase diagram of this model. The region with
$\gamma=\pi$ indicates a topologically nontrivial insulator and the
one with $\gamma=0$ indicates a trivial insulator. The gap overlap
region indicates a conductor.} \label{fig4}
\end{figure}

The analytical result has shown that $|M_z|=|2\sin{\theta}|$
separates the topologically non-trivial and trivial regimes with
$\gamma=\pi$ and $\gamma=0$. It is clear that $M_z= \pm
2\sin{\theta}$ are transition points, at which the energy gap is
closed and the Berry phase is not well defined because of the
degeneracy at the gap closing point. To establish the phase diagram,
we also need to ascertain the band overlap condition, under which
the system is a conductor. The two bands $E_{\pm}(k)$ overlap when
the two branches of the eigen-states satisfied
$min[E_{+}(k)]<max[E_{-}(k)]$.  From Eq.(\ref{eigenenergy}), we
observe $E_+(k)=-E_-(\pi-k)$. Hence, the two bands will overlap if
there exists a $k_0$ satisfied $E_{+}(k_0)<0$. Basing on these
conditions, we can determine phase boundaries and draw the phase
diagram of the half-filled system as displayed in Fig.\ref{fig4}.

\section{Symmetries}
Next we analyze the symmetries of the system described by
Eq.(\ref{H1}). The system has no time-reversal symmetry, as both
$t_s$ and $M_z$ terms break the time-reversal symmetry under the
time-reversal operation $\mathcal{T}=i\sigma_y \mathcal{K}$, with
$\mathcal{K}$ the complex conjugation operator. This Hamiltonian
also shows no conventional chiral symmetry defined as
$\mathcal{S}h(k)\mathcal{S}^{-1}=-h(k)$ in momentum space with
$\mathcal{S}$ representing a specific chiral operation.
However, we find $h(k)$ satisfies
\begin{equation}
\sigma_x h(k)\sigma_x=-h(\pi-k),
\end{equation}
which indicates the existence of a hidden chiral symmetry.
%hence this model shows the properties of the chiral unitary (AIII) class\cite{tenfold}.
In the case of finite number of lattice sites, we can rewrite the
real space Hamiltonian into the matrix form: $H=\hat{\Psi}^{\dagger}
h \hat{\Psi}$, with
$\hat{\Psi}^{\dagger}=(\hat{c}^{\dagger}_{1,\uparrow},\hat{c}^{\dagger}_{2,\uparrow},...
\hat{c}^{\dagger}_{L,\uparrow},\hat{c}^{\dagger}_{1,\downarrow},
\hat{c}^{\dagger}_{2,\downarrow},...\hat{c}^{\dagger}_{L,\downarrow})$.
We find that the hidden chiral operation can be realized by the
operator
\begin{eqnarray}
\mathcal{S} = \left(
\begin{array}{cc}
0 & \Upsilon\\
\Upsilon & 0
\end{array}\right),
\end{eqnarray}
%\begin{eqnarray}
%\mathcal{S}_{even}= \left(
%\begin{array}{ccccc}
%& & & & -i \\
%& & & i &  \\
%& & \iddots & &  \\
%& -i & & &  \\
%i & & & &
%\end{array}\right)
%\end{eqnarray} for even number of lattice sites $L$, and
%\begin{eqnarray}
%\mathcal{S}_{odd}= \left(
%\begin{array}{cc}
%I & 0 \\
%0 & -I
%\end{array}\right)\mathcal{S}_{even}
%\end{eqnarray}
%for the odd $L$ with $I$ the $L\times L$ identity matrix.
where $\Upsilon$ is a $L\times L$ matrix with $\Upsilon_{n,L-n+1}=(-1)^ni$ and other elements are 0.
This symmetry operator fulfills $\mathcal{S}^2=1$ for even $L$ and $\mathcal{S}^2=-1$ for odd $L$. Now it is
straightforward to verify that the Hamiltonian satisfies the chiral
symmetry \cite{odd_lattice}
\begin{equation}
\mathcal{S}h\mathcal{S}^{-1}=-h .
\end{equation}
Due to the existence of the chiral symmetry, for a eigenstate
$H\Phi=E\Phi$, there always exists another eigen-state
$\Phi'=\mathcal{S}\Phi$, which satisfies $H\Phi'=-E\Phi'$.

Besides the hidden chiral symmetry, we find that there also exists a
hidden symmetry under the operation defined as
$\mathcal{V}:\hat{c}_{n,\uparrow}\rightarrow
\hat{c}_{L-n,\uparrow},~\hat{c}_{n,\downarrow}\rightarrow
\hat{c}_{L-n,\downarrow},~i\rightarrow -i$. Such an operation is a
combination of the inversion and complex conjugation operations and
can be written explicitly as
\begin{eqnarray}
\mathcal{V}= \mathcal{K}\left(
\begin{array}{cc}
\Gamma & 0\\
0 & \Gamma
\end{array}\right),
\end{eqnarray}
where $\Gamma$ is a $L\times L$ matrix with all the elements on the
anti-diagonal are 1 and the other elements are 0. This operator also
has the property that $\mathcal{V}^2=1$, and we have
\begin{equation}
\mathcal{V}h\mathcal{V}^{-1}=h .
\end{equation}

The degenerate zero-mode edge states are protected by both the
hidden chiral symmetry and the hidden symmetry $\mathcal{V}$. The
hidden symmetry $\mathcal{V}$ indicate that if there is a edge state
$\Phi_L$ with eigen-energy $E_L$ localized at the left end, there
must be another edge state $\Phi_R=\mathcal{V}\Phi_L$ with $E_R=E_L$
localized at the right end. Meanwhile, the hidden chiral symmetry
indicates spectrum symmetric about $E=0$ and leads to $E_R=-E_L$.
Hence, the degenerated zero modes are protected by these symmetries
together. We note that a similar situation occurs in the SSH model,
for which the existence of zero modes is protected by both the
chiral symmetry and inversion symmetry \cite{Ryu}.

\section{Extended model with symmetry breaking} In this section, we study
an extended model by adding an on-site spin-flip term on the
Hamiltonian (\ref{H1}), which breaks the symmetries mentioned in the
last section. When either the hidden chiral symmetry or the hidden
symmetry $\mathcal{V}$ is broken, zero mode edge states are not
expected to appear. Nevertheless, we find a topologically nontrivial
phase can still exist even if the hidden chiral symmetry is broken.
Such a topological phase is protected by the hidden symmetry
$\mathcal{V}$ and can be still characterized by the Zak phase with
$\gamma=\pi$.

The Hamiltonian with the additional spin-flip term reads
\begin{eqnarray}
H = H_1 + M
\sum_{n=1}^L\hat{c}_{n,\uparrow}^{\dagger}\hat{c}_{n,\downarrow}+h.c.
\end{eqnarray}
In the limit case with $M_z=0$, the extended model can be mapped to
the Creutz ladder model \cite{Creutz} if we regard the two
components of fermions as two legs of the Creutz ladder. We note
that a scheme for the realization of the Creutz model in cold-atom
systems has been recently proposed \cite{optical-lattice1}. In the
momentum space, the Hamiltonian $h(k)$ is given by $h(k)=
d_0(k)I+d(k)\cdot\sigma$ with $d_0(k)=-2\cos{\theta}\cos{k}$,
$d_x(k)=2t_s\cos{k}+Re(M)$, $d_y(k)=-Im(M)$,
$d_z(k)=2\sin{\theta}\sin{k}+M_z$.
\begin{figure}
\includegraphics[width=0.9\linewidth]{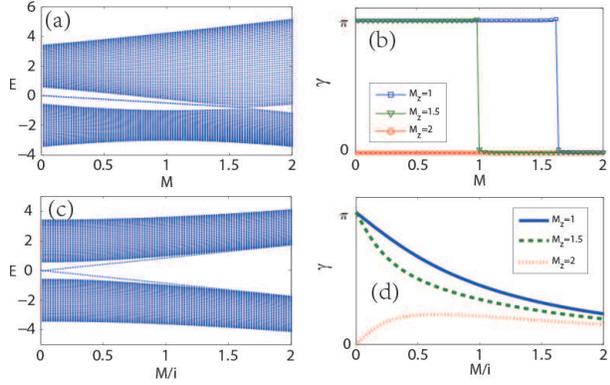}
\caption{(Color online) The energy spectrums and the Berry phase for
the extended model. (a) and (c) show the energy spectrum versus the
real and imaginary $M$, respectively, for the system with $M_z=1$,
$t_s=1$ and $\theta=\pi/3$ under the OBC; (b) and (d) show the
change of Berry phase as a function of real and imaginary $M$,
respectively, for the occupied Bloch band of systems with $t_s=1$,
$\theta=\pi/3$ and different $M_z$. (a) and (b) are for the real
$M$; (c) and (d) are for the imaginary $M$.} \label{fig5}
\end{figure}

For a real $M$, the on-site spin-flip term breaks the hidden chiral
symmetry. Consequently, for the system under the OBC, zero-mode edge
states no longer exist due to the breaking of chiral symmetry.
Nevertheless, the introduction of a real spin-flip term does not
break the hidden symmetry $\mathcal{V}$ of the system, and thus a
pair of degenerate edge states are still available. To see it
clearly, we display the energy spectra versus $M$ in
Fig.\ref{fig5}(a) for a system with $M_z=1$, $t_s=1$ and
$\theta=\pi/3$ under the OBC. As shown in the figure, the doubly
degenerate zero mode solutions at $M=0$ are not stable as the
eigenenergy deviates from $E=0$ in the presence of a  nonzero $M$
term. However, a pair of degenerate midgap states still exist in a
wide parameter regime with $M<M_{c}$, where $M_c$ is a transition
point with the gap closed. These midgap states are doubly degenerate
edge states and are protected by the hidden symmetry $\mathcal{V}$.
When $M>M_c$, a gap is reopened but no edge modes appear in the
regime of $M>M_{c}$. These results indicate that the transition
induced by changing $M$ is a topological transition. Similar to the
model discussed in the above sections, we can still characterize
this topological phase transition by the change of the Berry phase.
Following similar procedures in section II.C, we get the expression
of the Berry phase of the occupied Bloch band given by $
\gamma=\frac{\pi}{2} \left[ sgn \left(M_z- D \right )
 -sgn \left( M_z + D \right)\right]
$ for $M < 2 t_s$£¬ where
$D=2\sin{\theta}\frac{\sqrt{4t_s^2-M^2}}{2t_s}$. The result shows
that $\gamma=\pi$ in the regime of
$(\frac{M}{t_s})^2+(\frac{M_z}{\sin{\theta}})^2<4$, which agrees to
the numerical results in Fig.\ref{fig5}(b).

For the case with an imaginary $M$, the hidden chiral symmetry is
preserved, but the hidden symmetry $\mathcal{V}$ is broken. Fig.\ref{fig5}(c) and (d) shows the energy spectrum and the Berry phase
as a function of imaginary $M$. As the Berry phase is no longer a quantized number and the degenerate zero modes split into two branches, it clearly shows that topologically nontrivial states studied in this work are SPT states
protected by the hidden symmetry $\mathcal{V}$.

\section{Summary}
In summary, we have studied a topologically nontrivial fermion model
on a 1D lattice and demonstrated the existence of a conductor phase,
a trivial insulator phase and a topologically nontrivial insulator
phase for the half-filled system in different parameter regions. To
unveil nontrivial properties of the topological phase, we studied
both edge states of an open chain system and the Berry phase of the
corresponding bulk system in details, and identified that
topological phase can be characterized by either the existence of
doubly degenerate zero-mode edge states for the open system or a
quantized Berry phase for the bulk system. Together with the
condition of the band overlap, the phase diagram of the half-filled
system is also obtained. We also analyzed the symmetry of our model
and found that the topologically nontrivial phase is protected by
both a hidden chiral symmetry and a hidden symmetry described by the
combination of the inversion and complex conjugation operations.
Finally, we examined the hidden-symmetry-protected topological phase
by studying an extended model with symmetry breaking, and found that
a topologically nontrivial phase can still exist if only the hidden
chiral symmetry is broken. Such a topological phase can be
characterized by a quantized Berry phase or the existence of a pair
of degenerate non-zero-mode edge states, which is protected by the
hidden combination symmetry.

%S. C. would like to thank S. P. Kou and H. M. Guo for valuable
%discussions.
This work has been supported  by NSF of China under Grants No. 11374354,
No. 11174360, and No. 11121063.
% and by the Strategic Priority Research Program of
%the Chinese Academy of Sciences under Grant No. XDB07000000.


\begin{thebibliography}{99}
\bibitem{odd_lattice}This symmetry does not hold for odd $L$
sites with PBC. The symmetry matrix $\mathcal{S}$ works differently
for odd and even lattice sites, and the hopping term for even $L$
always connect an odd site and an even site. However, for odd $L$
with PBC, the hopping from one to another end connects two odd
sites, and breaks this symmetry. The numerical result also shows the
same property.
\end{thebibliography}
\end{document}